\documentclass[1p]{Article_website}
\usepackage{graphicx}
\usepackage{amsmath}
\usepackage{amssymb}
\usepackage{color,geometry}
\usepackage{psfrag}
\usepackage{pstricks}
\usepackage{fancyheadings}
\usepackage{enumerate}
\usepackage{nicefrac}
\definecolor{Green}{rgb}{0,0.7,0}
\definecolor{line1}{rgb}{0,0.447,0.741} 
\definecolor{line2}{rgb}{0.85,0.325,0.098} 
\definecolor{line3}{rgb}{0.466,0.674,0.188} 
\usepackage{wrapfig}
\usepackage{cleveref,hyperref}
\usepackage{wasysym}
\begin{document}
\title{Output-only parameter identification of a colored-noise-driven Van der Pol oscillator -- Thermoacoustic instabilities as an example}
\author{Giacomo  Bonciolini, Edouard  Boujo and Nicolas  Noiray}
\address{CAPS Laboratory, MAVT department ETH Z\"urich, 8092, Zurich, Switzerland}
\begin{abstract}
The problem of output-only parameter identification for nonlinear oscillators forced by colored noise is considered. 
In this context, it is often assumed that the forcing noise is white, since its actual spectral content is unknown.
%
The impact of this white noise forcing assumption upon parameter identification is quantitatively analyzed.
%
%
First, a Van der Pol oscillator forced by an Ornstein-Uhlenbeck process is  considered. 
Second, the practical case of thermoacoustic limit cycles in combustion chambers with turbulence-induced forcing is investigated.
It is shown that in both cases, the system parameters are accurately identified if time signals are appropriately band-pass filtered around the oscillator eigenfrequency.
\end{abstract}
\maketitle
\section{Introduction}
System identification (SI) is a long-standing problem that has fostered much research effort \cite{Ljung99, Pintelon12}. 
A wide variety of SI methods have been developed in different frameworks (control theory, machine learning, information theory), and tailored to  the specific situation at hand. 
In each case, the following questions, among others, must be considered to choose the adequate SI method: is it possible to apply a forcing to the system of interest and observe its response (input-output SI), or is it only possible to measure a given observable (output-only or ``blind'' SI)?
Is a model of the system already available, with parameters to be identified (parameter identification), or has the model itself to be uncovered (model identification)?
Does the system exhibit nonlinear  and/or transient behavior or can linear time invariance (LTI) be assumed?
Is the output corrupted by measurement noise?
Is the system itself subject to dynamic noise, i.e.  external stochastic forcing?
Is the system chaotic? 

Classical input-output SI techniques generally employ a state-space representation and estimate the parameters of a (postulated or physically derived) model by minimizing the error between the predicted and measured values of some output-based quantity, using e.g. maximum likelihood (ML), prediction error method (PEM), or least-squares (LS) \cite{Hamilton94, Shumway11, Pillonetto14, sovardi2016}. 
Popular model classes include auto-regressive / moving-average (AR, MA, ARMAX) models \cite{Liu10}, finite impulse response (FIR) models \cite{Polifke14}, output error models (OEM) \cite{Ding10} and Volterra series \cite{King16}.
If physical insight is lacking, SI can take care of selecting an adequate model among several candidates, although a careful trade-off between accuracy and simplicity is needed; this kind of Occam's razor principle is typically applied with probabilistic (Bayesian) approaches \cite{Beck10} or sparsity-promoting algorithms \cite{Chen14}.
Methods based on machine learning use kernels \cite{Pillonetto10} rather than postulating a model in the first place.

Output-only SI methods have to rely on partial information, either because the system cannot be arbitrarily driven, or because the input cannot be measured.
Standard tools include Kalman filters \cite{Kalman60}, synchronization methods \cite{Yu08}, modal identification \cite{Nagarajaiah09} and reduced-order modeling   \cite{Rowley17}.
Empirical dynamic modeling \cite{Ye15} allows for
model-free output-only SI.
As for input-output SI, sparse identification techniques are available for output-only SI \cite{Brunton2016}.

Rather than identifying a model or its parameters, some techniques allow the determination of a number of characteristics of a system: 
distinguish between its chaotic and stochastic nature \cite{Zunino12}, unveil time delays \cite{Zunino10} or discover  hidden patterns  \cite{Crutchfield12} based on information theory (e.g. entropy and complexity);
detect causality with convergent cross mapping \cite{Sugihara12};
analyze periodicity and intermittency in noisy signals using recurrence quantification analysis \cite{Eckmann87, Zbilut98, Suresha16}.

The presence of measurement noise and dynamic noise often complicates the task of SI, deteriorating both its accuracy when identifying parameters and its ability to select plausible models, even though state-space representations can explicitly account for noise.
See \cite{Reynders08, Zhang11, Kwasniok12}
for some efforts towards better noisy SI.
However, one can take advantage of the very presence of dynamic noise to extract information and perform output-only SI: inherent stochastic forcing drives the system away from its deterministic equilibrium trajectory and make it visit states that would not been visited otherwise.
As proposed in \cite{friedrich2011report}, these
enriched statistics  can then be processed to reconstruct the coefficients of the system's Langevin equation or corresponding Fokker-Planck equation \cite{Risken84} and identify the governing parameters.
\begin{figure*}[t]
\begin{center}
\includegraphics[width=150mm]{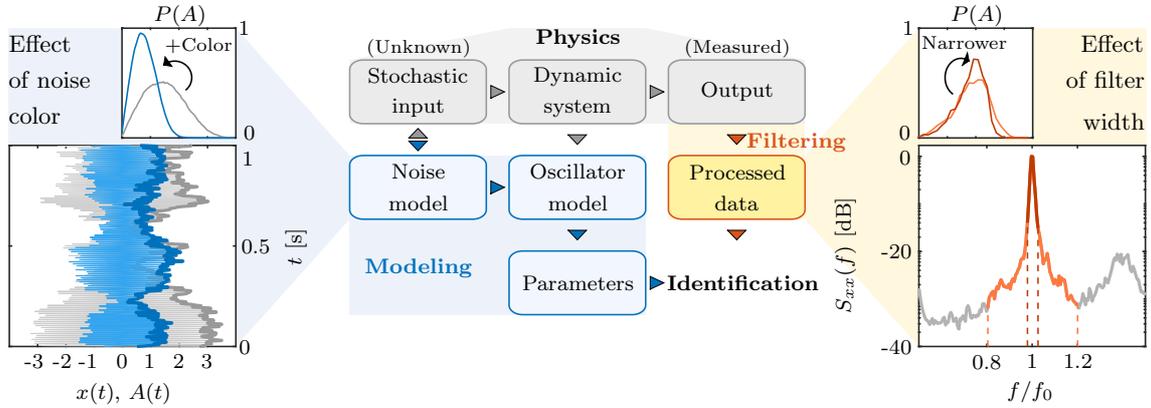}
\end{center}
\caption{Summary and context of the paper. Center:  stochastically forced dynamic system. In order to perform model-based output-only SI, models for the stochastic input and for the dynamic system (being in the present study a Van der Pol oscillator) are required. Left: effects of noise color on oscillator dynamics and statistics, with $x(t)$ being the system state and $A(t)$ its envelope (the energy is $\propto A^2$). Right: filtering the data to isolate the dynamics of interest can be needed. The corresponding filter bandwidth affects the statistics and dynamics of the data, which has to be accounted for in the parameter identification procedure.\label{fig:summary}}
\end{figure*}
In the present study this approach is adopted for output-only model-based SI for stochastically driven nonlinear oscillators: the parameters of a given analytical model are identified from the output signal of the system forced by a non-measurable random input.
Of course, in the case of linear harmonic oscillators, the system parameters (linear damping rate and resonance frequency) can be readily obtained, e.g. by estimating peak frequency and corresponding quality factor, which is not possible when nonlinearities are active.
In this context, accurate and robust output-only parameter identification requires:
\begin{enumerate}[(i)]
\item an adequate model of the system,
\item a model for the driving noise, 
\item an appropriate data pre-processing.
\end{enumerate}
\noindent{}These aspects are pictured in \cref{fig:summary}, where a summary of the present work is sketched.
In regards to point (i), the selected model for this work is a Van der Pol oscillator (henceforth ``VDP''), which is a canonical model used in many different disciplines such as electronics (since the pioneering work \cite{van1920theory}), biology and medicine \cite{van1928heart,jewett1998refinement,lucero2013modeling}, neurology \cite{fitzhugh1961impulses,nagumo1962active}, optics \cite{wirkus2002dynamics,barland2003experimental}, seismology \cite{cartwright1999dynamics}, sociology and economics \cite{glass1988clocks} or thermoacoustics dynamics in turbulent combustors \cite{noiray16symp}, the latter being the application discussed in more detail in the second part of the paper. The stochastic differential equation of a Van der Pol oscillator driven by additive noise reads:
\begin{equation} 
\label{VDPx}
\ddot{x}+\omega_0^{2}{x}=[2\nu-\kappa x^2]\dot{x} +\xi(t),
 \end{equation}
where $x$ represents the state of the system, $f_0=\omega_0/2\pi$ the natural oscillation frequency, $\nu$ the linear growth rate, $\kappa$ the saturation constant and $\xi(t)$ the additive driving noise.\\
\noindent{}Concerning point (ii), the simplest model for $\xi$ in \cref{VDPx} is the white noise because it greatly simplifies the analytical derivations. However, a real stochastic forcing is always ``colored", i.e. it always features a non-zero autocorrelation time and a non-constant spectral distribution. 
One can find a wide collection of studies where the color of the noise plays a fundamental role in the system dynamics, in topics such as economics, biology and mechanical configurations \cite{perello2002, qing2015, Sapsis08}, as well as in the specific case of oscillators \cite{masoliver1993,xu2011,spanos1978}. 
In the field of thermoacoustics, one can for instance refer to \cite{tony15} or \cite{waugh2011}, the latter investigating the effect of different types of noise on limit-cycle triggering. This suggests that it is essential to take the noise color into account in system identification.\\
In the first part of the present work, the widely used Ornstein-Uhlenbeck process is used as the driving source of the Van der Pol oscillator. Afterwards another type of noise is introduced for the specific case of thermoacoustic instabilities in turbulent combustors. In both cases, the associated system dynamics and statistics are scrutinised and the effect of noise color on parameter identification is addressed. The need of properly filtering the output data to reliably identify the parameters -- item (iii) in the aforementioned list -- is then discussed.
 \section{Van der Pol oscillator driven by Ornstein-Uhlenbeck noise}
 \label{OU}
 \subsection{Effect of colored noise on oscillations statistics}
 \label{OU1}
\begin{figure}[!t]
\begin{center}
\includegraphics[width=0.5\columnwidth]{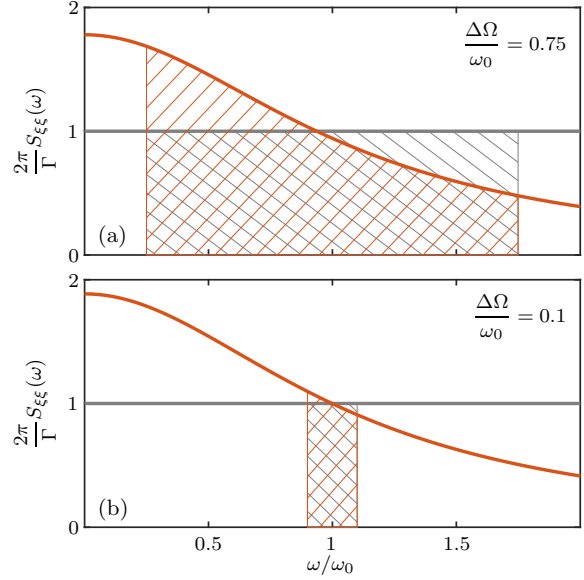}
\end{center}
\caption{Comparison between white noise (grey) and OU (red) noise power spectra, normalized by the white noise intensity $\Gamma$, for different iso-power semi-bandwidths $\Delta\Omega$. The power provided by the two types of noise is equal in the considered band (same area under the curve: note the linear scale). Note that $S_{\xi\xi}(\omega_0)=\Gamma_\text{e}/2\pi\neq \Gamma/2\pi$. \label{fig:OU_isopower}} 
\end{figure}

\begin{figure*}[!t]
\begin{center}
\includegraphics[width=\textwidth]{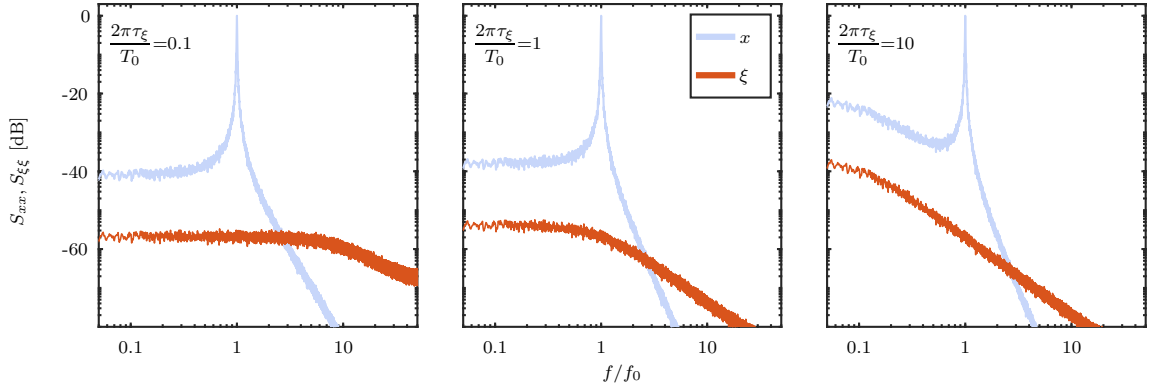}
\end{center}
\caption{Power spectra of the input $\xi$ and of the output $x$ for three different values of correlation time $\tau_\xi$, normalized by $T_0=1/f_0$. \label{fig:OU_fmax}}
\end{figure*}

\begin{figure}[t!]	
\begin{center}
\includegraphics[width=0.5\columnwidth]{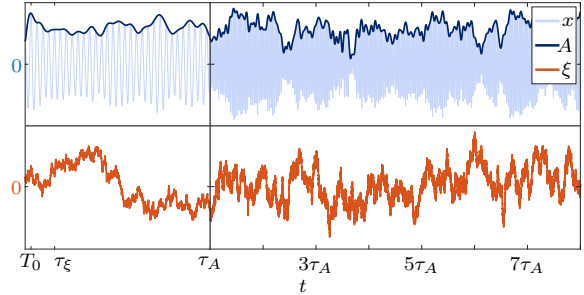}
\caption{Time scales involved in the stochastically forced oscillator: $\tau_\xi$ is the correlation time of the noise source $\xi$, $T_0=1/f_0$ is the oscillation period of $x$ and $\tau_A=\pi/|\nu|$ is envelope amplitude $A$ characteristic time scale. Note the two time different scales adopted for the two halves of the plot.}
\label{fig:tscale}
\end{center}
\end{figure}

\begin{figure*}
\begin{center}
\includegraphics[width=\textwidth]{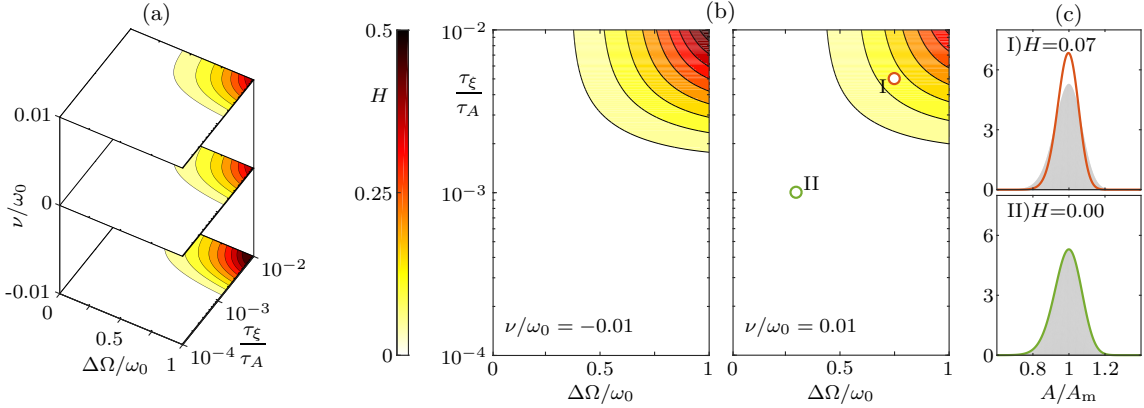}
\end{center}
\caption{Hellinger distance (\ref{H}), quantifying the difference between the PDFs of OU noise and white noise driven VDP oscillators. a) Different maps in the space $(\Delta\Omega,\tau_\xi,\nu)$. b) Detail of one linearly stable and one linearly unstable cases. c) PDFs comparison of two points in the map. The correlation time of the noise is normalized on the correlation time of the pressure amplitude $\tau_A=\pi/|\nu|$. The linear growth rate $\nu$ is normalized on the oscillator's angular frequency $\omega_0$.\label{fig:OU_H}}
\end{figure*}

\noindent{}\noindent In this section, the noise that drives the Van der Pol oscillator is generated by an Ornstein-Uhlenbeck (OU) process. It is widely used in various contexts to account for finite correlation time effects of a stochastic forcing. 
One therefore considers that $\xi$ in \cref{VDPx} satisfies the following Langevin equation:
\begin{equation}
\label{OU_de}
\dot{\xi}(t)=-\dfrac{1}{\tau_\xi}\xi(t)+\dfrac{\sqrt{\gamma}}{\tau_\xi}\zeta(t),
\end{equation}
where $\zeta$ is a unit-variance Gaussian white noise of intensity $\Gamma$, $\gamma$ is a constant coefficient, which will be used later in the paper to adjust the power of the noise $\xi$, and $\tau_\xi$ denotes its characteristic time constant.
In the frequency domain, the OU process $\widehat{\xi}$ results from filtering $\widehat{\zeta}$ with the following transfer function
\begin{equation}
\label{OU_transfer_function_noise}
H(s)=\frac{\widehat{\xi}(s)}{\widehat{\zeta}(s)}=\frac{\sqrt{\gamma}}{1+\tau_\xi s},
\end{equation}
where $s=i\omega$ is the Laplace variable. The power spectrum of $\xi$ is given by
\begin{equation}
\label{OU_power_spectrum_noise}
S_{\xi\xi}(\omega)=|H|^2S_{\zeta\zeta}=\frac{\Gamma}{2\pi}\frac{\gamma}{1+\omega^{2}\tau_\xi^{2}},
\end{equation}
It is useful to define the quantity
\begin{equation}
\label{Gamma_ou}
\Gamma_\text{e}=2\pi S_{\xi\xi}(\omega_0)=\Gamma\frac{\gamma}{1+\omega_0^{2}\tau_\xi^{2}},
\end{equation}
which is the power spectral density of $\xi$ at the oscillator eigenfrequency, referred to as ``effective OU noise intensity" in the remainder of the paper.\\
Considering that the target of this study is to quantitatively compare white and colored noise forcing on the oscillator, it is necessary to set a criterion regarding the input power. It is convenient to adjust the intensity of $\xi$ by using the coefficient $\gamma$ such that the powers provided by $\xi$ and by a white noise of intensity $\Gamma$ in a band $[\omega_{1};\omega_{2}]$ are equal, i.e. $\int_{\omega_1}^{\omega_2}S_{\xi\xi}d\omega=\int_{\omega_1}^{\omega_2}{\Gamma}/{2\pi}d\omega$, which yields:
\begin{equation}
\label{OU_intensity_noise}
\gamma=\frac{\tau_\xi(\omega_2-\omega_1)}{\text{atan}(\omega_{2}\tau_\xi)-\text{atan}(\omega_{1}\tau_\xi)}.
\end{equation}
A sensible choice is to define this ``iso-power band'' around the oscillator resonance frequency $\omega_0$: $[\omega_{1};\omega_{2}]=[\omega_{0}-\Delta\Omega;\omega_{0}+\Delta\Omega]$.
Henceforth, $\Delta\Omega$ can vary between 0 (band degenerating in the single angular frequency $\omega_{0}$) and $\omega_0$ (band $[0;2\omega_0]$). The frequency range $\Delta\Omega$ will be referred to as ``iso-power semi-bandwidth''. One can see in \cref{fig:OU_isopower} how this parameter affects the forcing noise power spectrum. 
The parameter $\tau_\xi$ is a direct measure of how much ``colored'' the noise is: the shorter $\tau_\xi$, the closer to a white noise $\xi$ is. 
As $\tau_\xi$  goes to zero, the cut-off frequency goes to infinity, leading to a constant power spectrum, i.e. a white noise source. This is illustrated in \cref{fig:OU_fmax} (red spectra), together with the fact that the power spectral density of the oscillator response (blue spectra) is accordingly affected. Note that in the limit $\tau_\xi\rightarrow0$, one gets $\gamma\rightarrow1$ and $S_{\xi\xi}(\omega)\rightarrow{\Gamma}/{2\pi}=S_{\zeta\zeta}(\omega)$. The characteristic time $\tau_\xi$ is the noise correlation time, obtained via the autocorrelation function of $\xi$:
\begin{equation}
\label{OU_autocorrelation_function}
k_{\xi\xi}(t)=\Gamma\frac{\gamma}{2\tau_\xi}e^{-\frac{t}{\tau_\xi}}, \,\,\,\,\,\, \tau_\xi=\frac{1}{k_{\xi\xi}(0)}\int_{0}^{\infty}|k_{\xi\xi}(t)|dt.
\end{equation}
%
\noindent Such OU process is now considered as being the driving force of the Van der Pol oscillator given by \cref{VDPx}. It is convenient to investigate the system in terms of its slowly-varying amplitude and phase dynamics with $x(t)\approx A(t)\cos{[\omega t + \varphi(t)]}=A(t)\cos{\phi(t)}$. This coordinate change is legitimate provided that $\nu\ll\omega_0$. Performing deterministic and stochastic averaging \cite{stratonovich1967} yields the following stochastic differential equation for the amplitude $A$:
\begin{equation}
\label{OU_A_dot_vdp_avg} 
\dot{A}=A\left(\nu-\frac{\kappa}{8}A^{2}\right)+\frac{\Gamma_\text{e}}{4\omega_0^2 A}+\mu(t),\hspace{5mm}
\text{with}\quad\ \langle\mu\mu_\tau\rangle=\frac{\delta(\tau)\pi S_{\xi\xi}(\omega_0)}{\omega_0^2}=\frac{\Gamma_\text{e}}{2\omega_0^2}\delta(\tau).
\end{equation}
\noindent{} It is important to underline that the averaging procedure is valid only if $\tau_{\xi}\ll\tau_A$, where the amplitude correlation time $\tau_A$ is related to the system growth rate by $\tau_A \simeq \pi/|\nu|$ (see \cite{lax_1967,lax_2006,noiray16}). One can refer to  \cref{fig:tscale} where the important time scales of the considered system  are presented. It is also interesting to compare \cref{OU_A_dot_vdp_avg} to its white-noise-driven oscillator counterpart
\begin{equation}
\dot{A}=A\left(\nu-\frac{\kappa}{8}A^{2}\right)+\frac{\Gamma}{4\omega_{0}^{2}A}+\mu(t), \hspace{5mm}
\text{with}\quad \langle\mu\mu_\tau\rangle=\frac{\delta(\tau)\pi S_{\zeta\zeta}(\omega_0)}{\omega_0^2}=\frac{\Gamma}{2\omega_0^2}\delta(\tau).
\label{Aw_dot_vdp_avg} 
\end{equation}
The two equations only differs by the fact that $\Gamma_\text{e}$ substitutes $\Gamma$. In the limit $\tau_\xi\rightarrow0$, $\Gamma_\text{e}\rightarrow\Gamma$, and \cref{OU_A_dot_vdp_avg} tends to \cref{Aw_dot_vdp_avg}. 
Considering the Fokker-Planck equation associated with \cref{OU_A_dot_vdp_avg}, one can derive the stationary probability distribution (PDF) for the amplitude of the  VDP oscillator driven by an OU noise:
\begin{equation}
P_{\mathrm{ou}}(A)=\mathcal{N}_{\mathrm{ou}}A\exp{\left[\frac{4\omega_{0}^{2}}{\Gamma_\text{e}}\left(\frac{\nu A^{2}}{2}-\frac{\kappa A^{4}}{32}\right)\right]},
\label{OU_PAc}
\end{equation}
and for the white-noise driven VDP oscillator:
\begin{equation}
P_\text{w}(A)=\mathcal{N}_\text{w}A\exp{\left[\frac{4\omega_{0}^{2}}{\Gamma}\left(\frac{\nu A^{2}}{2}-\frac{\kappa A^{4}}{32}\right)\right]},
\label{PAw}
\end{equation}
where $\mathcal{N}_{\mathrm{ou}}$ and $\mathcal{N}_{\mathrm{w}}$ are two normalization constants such that $\int_{0}^{\infty}P(A)dA=1$.
\begin{figure}
\begin{center}
\includegraphics[width=0.5\columnwidth]{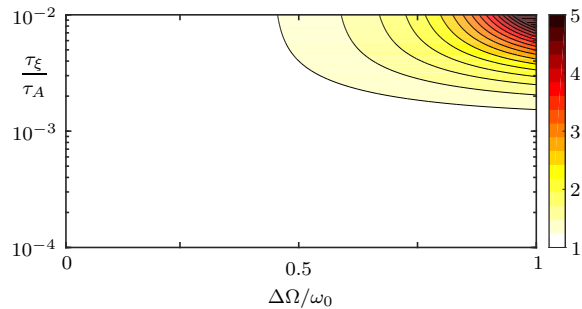}
\end{center}
\caption{Map of the coefficient ${\Gamma/\Gamma_\text{e}=(1+\omega_{0}^{2}\tau_\xi^{2})}/{\gamma}$. The closer this is to one, the closer the analytical expressions for $P_{\mathrm{ou}}$ and $P_{\mathrm{w}}$ are.\label{fig:OU_fact_map}}
\end{figure}
Apart from the normalization constants, $P_{\mathrm{ou}}$ and $P_{\mathrm{w}}$ differ by the factor $\Gamma/\Gamma_\text{e}={(1+\omega_{0}^{2}\tau_\xi^{2})}/{\gamma}$  in the exponential, which is depicted in \cref{fig:OU_fact_map}.
\begin{figure}[t!]
\begin{center}
\includegraphics[width=0.5\columnwidth]{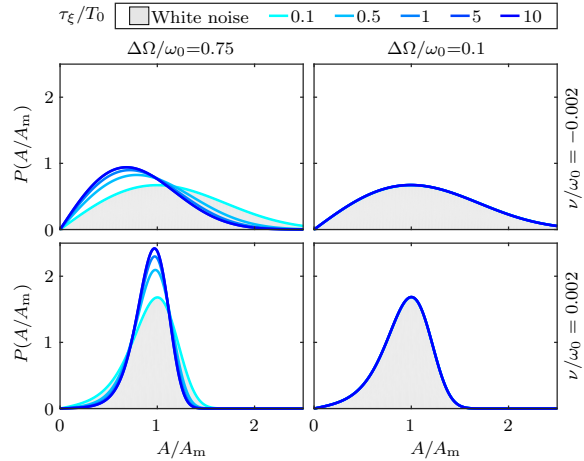}
\end{center}
\caption{\color{black} Probability density function for two different linear growth rates $\nu$, two different iso-power semi-bandwidth $\Delta\Omega$, and five different adimensional correlation times $\tau_\xi/T_0$ of the driving noise (where $T_0=2\pi/\omega_0$ is the oscillation period).  Shaded area and solid lines respectively correspond to the PDFs of the VDP driven by white noise ($P_{\mathrm{w}}$ given in \cref{PAw}) and to the VDP driven by the OU noise ($P_{\mathrm{ou}}$ given in \cref{OU_PAc})  for the same parameters $\nu$, $\kappa$, $\omega_0$ and $\Gamma$. The amplitude $A$ is normalized by $A_{\mathrm{m}}$, which is the amplitude of the maximum of $P_{\mathrm{w}}$. \label{fig:OU_vdp_pdf}}
\end{figure}\\
\Cref{fig:OU_vdp_pdf} compares the amplitude PDFs of the oscillator driven by white noise (shaded area) and by the colored noise (solid lines) for the same system parameters $\nu$, $\kappa$, $\omega_0$ and $\Gamma$. 
Two different iso-power semi-bandwidths $\Delta\Omega$ (columns), which were already considered in  \cref{fig:OU_isopower},  as well as two different values of the linear growth rates $\nu$ (rows) are considered. In the case of a wide iso-power band, one can observe that  $P_{\mathrm{ou}}$ significantly deviates from $P_{\mathrm{w}}$ when $\tau_\xi$ increases. One can note that for large enough $\tau_\xi$ and for $\Delta\Omega<\omega_0$, $P_{\mathrm{ou}}$ tends to a limit case distribution\footnote{It can be proven that $\lim_{{\tau_\xi}\to\infty} (1+\omega_{0}^{2}\tau_\xi^{2})/\gamma=\omega_0^2/(\omega_0^2-\Delta\Omega^2)$, so except for the case $\Delta\Omega=\omega_0$, this limit is finite and the $P_{\mathrm{ou}}$ asymptotically tends to a limit PDF. Remember that $\tau_\xi\ll \tau_A$ must anyway hold to have a valid derivation of the equations.}. On the other hand, no significant difference among the PDFs can be noticed when $\Delta\Omega/\omega_0=0.1$.\\
\noindent{}To obtain a quantitative measure of the difference between the two PDFs, one can make use of the Hellinger distance:
\begin{equation}
\label{H}
H=\sqrt{1-B},
\end{equation}
where $ B=\int_{-\infty}^{+\infty}\sqrt{p(x)q(x)}\mathrm{d}x $ is the Bhattacharyya coefficient.
The Hellinger distance $H$ is a statistic quantity that measures the difference between two PDFs of the same random variable $p(x)$ and $q(x)$, and ranges from 0 when $p(x)=q(x)$, to 1 when they do not overlap. 
In the following, $H$ is computed to compare $P_\text{w}$ and $P_\text{ou}$ in a systematic way for different points $(\Delta\Omega,\tau_\xi,\nu)$ of the space of iso-power semi-bandwidth, correlation time and growth rate. The results are presented as colormaps in \cref{fig:OU_H}.
\noindent{}The linear growth rate $\nu$ has a minor effect: all the maps in \cref{fig:OU_H}.a are similar, but $H$ is slightly higher when $\nu<0$, due to the shift of the amplitude of maximum probability $A_{\mathrm{m}}$ observed in this case (see again \cref{fig:OU_vdp_pdf}). Focusing on the other two parameters in \cref{fig:OU_H}.b, $H$ is large in the upper-right corner of the map, i.e. for high values of $\Delta\Omega$ and $\tau_\xi$. The influence of $\tau_\xi$ is intrinsically related to the noise color:  as discussed earlier, the shorter $\tau_\xi$, the closer is $\xi$ to a white noise. That is why the region of match between $P_{\mathrm{c}}$ and $P_{\mathrm{w}}$ is wider for short correlation times. In case of a long $\tau_\xi$, the bandwidth $\Delta\Omega$ has a strong influence, leading for large values to a significant difference between $P_{\mathrm{c}}$ and $P_{\mathrm{w}}$. 
A large $\Delta\Omega$ means that the equality of power between white noise and colored noise is set in a wide band around the oscillator eigenfrequency. If $\tau_\xi$ is long enough to let the oscillator frequency $f_0$ fall in the decaying part of $S_{\xi\xi}$, the power spectral density of the two forcing noise is sensibly different around $f_0$ (see again \cref{fig:OU_isopower}), and the response of the system significantly changes.	
\subsection{Parameter identification and white-noise assumption}
\label{OU2}
\noindent{}In this section, the influence of the finite correlation time $\tau_\xi$ of the driving OU noise upon parameter identification strategies is investigated. 
The problem is the following: the noise driving the oscillator is never white in practice. Therefore, the use of a white noise driven oscillator model as a base for parameter identification can be brought into question.\\
One alternative would be to adopt a model featuring a noise source with finite correlation time as exemplified in the previous section with the OU process. However, this would not make any difference if the adopted SI method relies on the statistics of the signal. In fact, looking at \cref{OU_PAc}, one can see that the analytical expression for $P_\text{ou}$ produces self-similar probability distributions. In other words, different combinations of $\Gamma$ and $\tau_\xi$, lead to the same output amplitude statistics. 
However, if one is only interested in identifying  the linear growth rate $\nu$ and the saturation coefficient $\kappa$ one should presumably be able to use a white noise driven VDP model as a basis for the SI.
Still, it has to be verified if the presence of non-zero autocorrelation time $\tau_\xi$ can affect the identification process: even though the amplitude PDFs of the two models are the same, the output time traces and spectra are different, especially for some combinations of parameters.\\
\noindent{}To verify the possibility of achieving a robust parameter identification of the linear growth rate $\nu$ and the saturation coefficient $\kappa$ using a white noise approximation, the following test is performed. A Van der Pol oscillator (see \cref{VDPx}), having the true parameters $\nu=\nu_\text{t}$ and $\kappa=\kappa_\text{t}$ and forced with an OU noise of intensity $\Gamma_\text{t}$ and correlation time $\tau_\xi$, is simulated in Simulink\textsuperscript{\textregistered}, and then the slowly-varying envelope $A(t)$ and phase $\varphi(t)$ of the output signal $x(t)$ are extracted.
%
\begin{figure*}
\begin{center}
{\includegraphics[width=\textwidth]{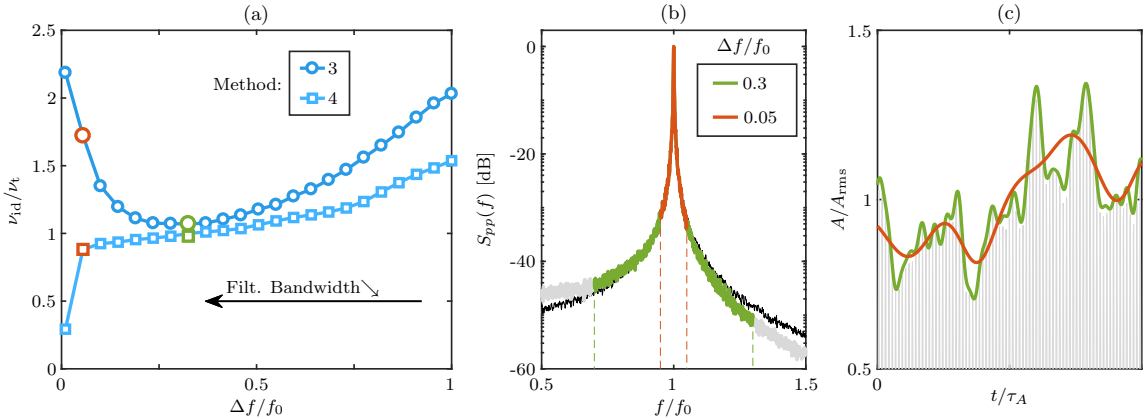}}
\end{center}
\caption{a) Identified growth rate as a function of the filter bandwidth. b) Power spectrum of the signal (grey), and two filter windows (color highlight) for the VDP driven by OU noise. The thin black line is the spectrum of a VDP driven by a white noise of intensity $\Gamma_\text{e}$. c) Amplitude time traces (red and green) obtained from the two proposed filters, superimposed to the unfiltered signal (grey). In panel a) the corresponding points are highlighted.\label{fig:OU_filt_width}}
\end{figure*}
A parameter identification using the white noise driven model is then attempted, making use of the approaches 3 and 4 proposed in \cite{noiray13}. They consist in finding the optimum parameters $\nu$, $\kappa$ and $\Gamma$ giving the best fit of $P(A)$ and $P(A\dot{\varphi})$ for method 3, and of the drift and diffusion coefficients of the Fokker-Planck equation for method 4. However, the identified parameters significantly differ from the actual values: $\nu_\text{id}=2.1\nu_\text{t}$, $\kappa_\text{id}=1.9\kappa_\text{t}$, $\Gamma_\text{id}=1.7\Gamma_\text{t}$ with approach 3 and $\nu_\text{id}=1.5\nu_\text{t}$, $\kappa_\text{id}=1.6\kappa_\text{t}$, $\Gamma_\text{id}=1.5\Gamma_\text{t}$  using the approach 4.\\

\noindent{}As will become apparent, the parameter identification failed because of the lack of pre-processing of the data. 
In fact it is wrong to assume that the measured output spectrum $S_{xx}(\omega)$ can be generated by an equivalent white noise source: the actual driving noise spectral power distribution $S_{\xi\xi}(\omega)$ leaves some peculiar signature in $S_{xx}(\omega)$. However, it is indeed possible to reproduce over a limited band around the oscillator frequency the actual output of the colored noise driven VDP with a white noise forcing, because $S_{\xi\xi}$ is a smooth function of frequency. This is exemplified in \cref{fig:OU_filt_width}.b, where one can see the spectrum of a colored noise driven VDP (thick grey line), overlaying the one of a VDP driven by a white noise of intensity $\Gamma_\text{e}$ (thin black line).\\

\noindent{}The next attempt is, therefore, to bandpass filter the signal obtained from the simulation in the band $f_0\pm\Delta f$, using a $\Delta f$ progressively narrower\footnote{Note that $\Delta f$ is not related to $\Delta\Omega$: the first is the filter semi-width adopted to pre-process the data for parameter identification, the second is a semi-bandwidth arbitrarily chosen to define the driving noise intensity.}. The obtained identification of $\nu$ is presented in \cref{fig:OU_filt_width}.a as a function of $\Delta f$. If $\Delta f=f_0$, the identified parameters values are close to the ones obtained using the unfiltered data. Decreasing $\Delta f$, the identified growth rate $\nu_\text{id}$ converges to the actual one $\nu_\text{t}$ for $\Delta f/ f_0=0.3$. The same trend is found for the saturation constant $\kappa$. This indicates that it is necessary to filter the data around the frequency of interest in order to perform a reliable model-based output-only parameter identification.\\

\noindent{}One might be tempted to reduce further the filter bandwidth, in order to decrease even more the driving noise modeling inaccuracy. However, one can see that below $\Delta f=0.05$ the estimated $\nu$ again deviates from the actual one. This fact is explained through the other two panels of \cref{fig:OU_filt_width}. In the panel b, the spectrum of the signal generated by the simulation of the OU noise driven VDP oscillator is presented, together with two different filter widths. The corresponding filtered time traces of the oscillation amplitude, used as data for the parameter identification, are plotted in panel c, superimposed to the unfiltered oscillator signal (grey). One can observe that if a too narrow band is considered, the signal is altered and substantially deviates from the original: the amplitude time trace follows the general trend, but does not capture anymore the high frequency content. This affects the statistics and dynamics of the data and, therefore, the outcome of the parameter identification. Hence, one must refrain from filtering too much the signal, to preserve the core information of the original signal.\\

\begin{figure}[t!]
\begin{center}
\includegraphics[width=0.5\columnwidth]{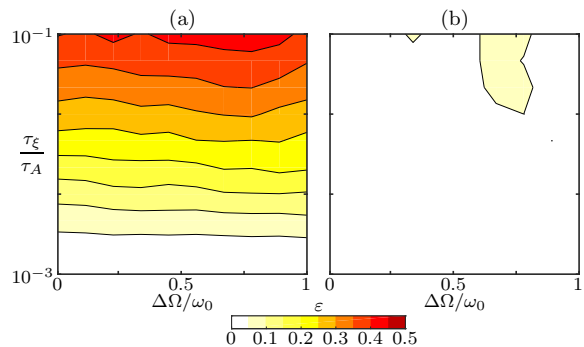}
\end{center}
\caption{Map of OU noise driven oscillator identification, for different noise correlation time $\tau_\xi$ and iso-power semi-bandwidth $\Delta\Omega$. a) Identification using the unfiltered data. b) Identification using signals filtered in the band $[f_0(1\pm0.5)]$. The identification result is given as relative error: $\varepsilon=|\nu_\text{t}-\nu_\text{id}|/\nu_\text{t}$.\label{fig:OU_ID_map}}
\end{figure}

\noindent{}In the next step, the parameter identification is performed for different colored noise parameters, to ensure that an adequate filtering is the means of achieving a reliable identification. In \cref{fig:OU_ID_map} the results of this test are presented. Each panel includes the identification result (method 4 in \cite{noiray13} is adopted) of 100 different simulations of the system, each corresponding to a different combination of noise parameters $\Delta\Omega$ and $\tau_\xi$. The identification inaccuracy is given in terms of relative error $\varepsilon=|\nu_\text{t}-\nu_\text{id}|/\nu_\text{t}$. In the left column, the identification results when using the raw data are presented. The iso-power semi-bandwidth $\Delta\Omega$ does not noticeably affect this error, as it just changes the value of $\Gamma_\text{e}$ to be identified. The noise correlation time has a dramatic impact on the identification error for long correlation times. However, the error vanishes if $\tau_\xi$ is very short, as in this case the driving noise gets closer a white one.
In the right column of \cref{fig:OU_ID_map}, the same signals are bandpass filtered in a band $[f_0(1\pm0.5)]$ before the parameter identification is run. One can notice how the identification is considerably enhanced, leading to very accurate results regardless of the parameters of the noise source. This result consolidates the confidence on the output-only parameter identification methods, as even without knowing the noise parameters $\Gamma$, $\Delta\Omega$ and $\tau_\xi$ it is possible to obtain the correct oscillator parameters $\nu$ and $\kappa$ by just applying an adequate filter to the output signal. 
\\ 
\noindent{}Summing up, it can be stated that, for a OU noise driven VDP oscillator, the parameter identification based on a white noise approximation will accurately estimate the linear growth rate $\nu$ and the saturation constant $\kappa$ if the signal is filtered before the analysis. The filtering bandwidth has to be:
\begin{itemize}
\item narrow enough, to have a satisfactory approximation of the real noise with a white one over the considered band,
\item not too narrow, to preserve the amplitude dynamics of the signal.
\end{itemize}
A sensible strategy is to use a progressively narrower filter for the data pre-processing, and repeat the identification process until the obtained parameters reach a plateau.\\
In the next part of this work, the study will be carried out using a different type of noise source, which is also closer to the actual stochastic forcing characteristic of thermoacoustic systems.
\section{Thermoacoustic instabilities: Modeling}
\label{Thermoac}
\subsection{Practical context}
\label{intro_TA}
\noindent{}In gas turbine, aeronautics and aerospace applications, the race for more efficient, less polluting, more fuel- and operation-flexible systems is ongoing, towed by customers needs and environmental regulations \cite{lieuwen2012book}. The thermoacoustic instabilities taking place in the combustion chambers of these engines constitute a major difficulty to overcome  \cite{poinsot2016}, because their resulting high amplitude acoustic levels induce high cycle fatigue of the combustor components and reduce their lifetime. The mechanisms ruling the constructive interaction between flames and acoustic modes are complex and the occurrence of these instabilities at a given engine operating point is hard to predict.\\
{\noindent}Therefore, the development of reliable predictive methods is of primary importance. Currently, brute force Large Eddy Simulations cannot be routinely used in a combustor design optimisation context due to their prohibitive computational costs. Therefore, a significant portion of the research efforts concentrate on the development of Helmholtz solvers and low-order thermoacoustic network models that are combined with experiments or computationally-cheaper numerical simulations \cite{schuermans2010, han2015,silva2017,nicoud2007,bourgouin2015,oberleithner2015,campa2014,schmid2013,ghirardo2015}.\\

\noindent{}In parallel, it is also important to establish robust system identification methods in order to validate the aforementioned linear-stability prediction tools. It has been recently shown that  thermoacoustic linear growth rates can be extracted from limit cycle dynamic pressure data recorded in real systems \cite{noiray13,noiray2013dynamic,noiray16,noiray16symp}, and compared to the ones obtained using predictive thermoacoustic methods. Such network model validation is performed in \cite{bothien2015analysis}.\\
In the context of the present work, this section deals with output-only parameter identification methods applied to thermoacoustic systems, where the measurable output is the acoustic pressure at one location in the combustion chamber while the unknown input is the stochastic forcing resulting from the intense turbulence in the combustor. This last contribution is often modelled as an additive forcing, and assumed to be a white noise. In reality, this noise is not delta-correlated as explained in section \ref{Combustion noise}. Therefore, in section \ref{VDP_model} a more accurate model of the actual noise is introduced and the equations for the Van der Pol forced by this specific noise source are derived. The impact of the selected model on the effectiveness of the parameter identification is afterward scrutinised in section \ref{Validity of the white noise approximation}.\\
Regarding the system modeling, a single thermoacoustic mode description is often adopted in order to keep the number of system parameters to be identified to a minimum. This allows the use of a single oscillator as a model of the system. However the raw data, i.e. the acoustic pressure at a given location in the combustion chamber, result from the superposition of the contributions from all the combustor eigenmodes. Consequently, it cannot be directly treated and requires pre-processing to isolate the information corresponding to the single eigenmode considered for parameter identification. This can be done by bandpass filtering the data \cite{noiray16symp} or by performing a modal projection if several simultaneous records at different locations in the chamber are available \cite{noiray2013dynamic}. These data manipulations can, however, change the outcome of output-only parameter identification methods, because the signal and its statistics can be sensibly altered. This problem is considered  in section \ref{Filter size for system identification}.
\subsection{Colored random excitation}
\label{Combustion noise}

\begin{figure}	
\begin{center}
\includegraphics[width=0.5\columnwidth]{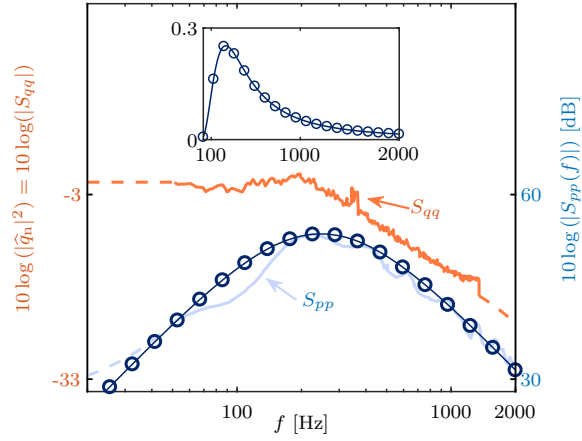}
\caption{Example of power spectra ($S_{qq}$ and $S_{pp}$) of turbulence-induced heat release rate fluctuations $\widehat{q}_{\mathrm{n}}$  and combustion noise of a flame radiating sound in the free field (adapted  from reference \cite{rajaram2009}). $S_{pp}$ can be approximated with a bandpass model (-$\Circle$-), plotted also in the inset, in linear scale. \label{fig:exp_Spp_Sqq_open}}
\end{center}
\end{figure}
\begin{figure*}	
\begin{center}
\includegraphics[width=\textwidth]{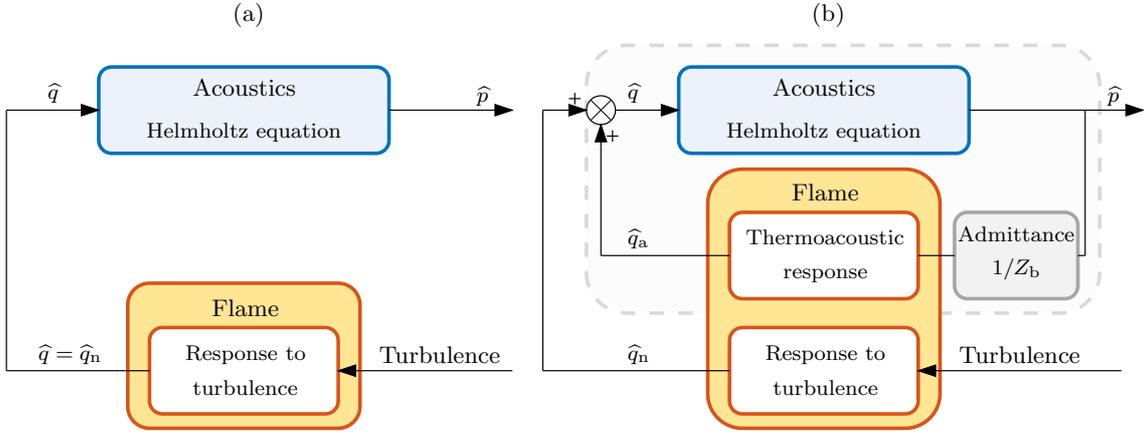}
\caption{Block diagram for the sound field generated by turbulent flames in open and closed environments. The open loop configuration (a) corresponds to open flames radiating noise in the free field. When the flame is enclosed (b), a thermoacoustic feedback operates and the acoustic block is fed by the total heat release rate fluctuations $\widehat{q}=\widehat{q}_{\mathrm{a}}+\widehat{q}_{\mathrm{n}}$, where $\widehat{q}_{\mathrm{a}}$ and $\widehat{q}_{\mathrm{n}}$ respectively stand for the acoustically- and turbulence-induced components.\label{fig:network}}
\end{center}
\end{figure*}
\begin{figure}[t!]
\begin{center}
\includegraphics[width=\columnwidth]{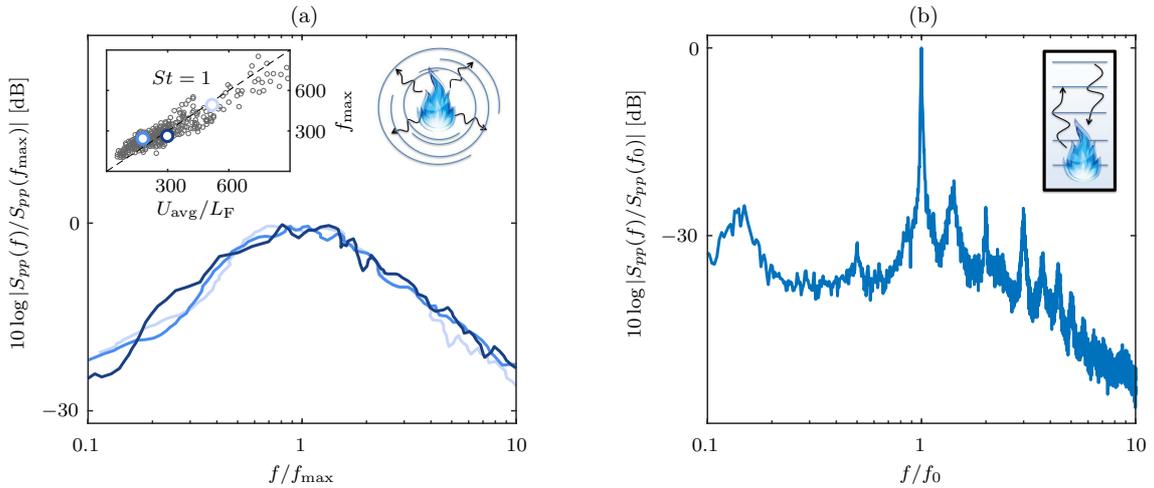}
\end{center}
\caption{a) Example of normalized combustion noise spectra measured for different open flames configurations (adapted from \cite{rajaram2009}). In the inset, the frequency of the spectrum maximum $f_{\mathrm{max}}$ is given as a function of the flow characteristics for a large set of operating conditions (see main text for definitions). b) Typical acoustic pressure spectrum recorded in a combustion chamber.\label{fig:exp_Spp_engine}}
\end{figure}
\noindent{}In thermoacoustics, the acoustic pressure satisfies the Helmholtz equation  with heat release rate source in the volume of the domain and the impedance conditions on boundaries:
\begin{equation}
\label{helmholtz}
\nabla^{2}\widehat{p}(s,x)-\left(\frac{s}{c}\right)^2\widehat{p}(s,x)=-s\frac{(\gamma-1)}{c^{2}}\widehat{q}(s,x) \hspace{5mm} \text{in the domain},
\end{equation}	
\begin{equation}
\label{helmholtz_BC}
\frac{\widehat{p}(s,x)}{\mathbf{\widehat{u}}(s,x)\cdot \mathbf{n}}=Z(s,x) \,\,\,\,\, \text{on boundaries},
\end{equation}
where $\widehat{p}$ and $\widehat{u}$ are the acoustic pressure and velocity fluctuations, $s$ the Laplace variable, $x$ the position, $c$ the local speed of sound, $\gamma$ the specific heat ratio, $\widehat{q}$ the heat release rate fluctuation, $\mathbf{n}$ the outward normal to the boundary and $Z$ the acoustic impedance. This equation stands if the Mach number is low. If the flame is placed in an open environment, waves generated by the reaction zone are radiated away without reflections. In reference \cite{hirsch2007}, the radiated sound field in this situation is modelled as function of the turbulence-induced heat release rate fluctuation and compared to experimental data. The formal solution of \cref{helmholtz} for a fluctuating heat release rate source in an open environment is:
\begin{equation}
\label{helmholtz_sol_OF}
\widehat{p}(s,x)=s\frac{(\gamma-1)}{4\pi r c^{2}}\int_{V_{f}}\widehat{q}(s,y)e^{\frac{s}{c}|x-y|}d^3y,
\end{equation}	
where $x$ is the observer position in the far field, $r\approx|x|$ is the distance of the observer from the flame. This equation is valid when the flame brush, which extends over the volume $V_f$, is compact with respect to the considered acoustic wavelength.
An example of the far-field acoustic power spectral density  $S_{pp}$ in such configuration, i.e. the so-called \emph{combustion noise} \cite{strahle78},  is given in \cref{fig:exp_Spp_Sqq_open}, together with integrated heat release oscillation power spectrum $S_{qq}$. In this situation, the heat release rate fluctuations $\widehat{q}$ generating the sound field are only due to the non-coherent turbulent component $\widehat{q}_\text{n}$ (see \cref{fig:network}.a).\\ 
\noindent{}The combustion noise spectrum $S_{pp}$ features a maximum at frequency $f_{\mathrm{max}}$ and a bandpass signature, in contrast with the low-pass character of $S_{qq}$, having $f_{\mathrm{max}}$ as cut-off frequency.  The two spectra are related to each other by \cref{helmholtz_sol_OF}, which is the topic of e.g. \cite{rajaram2009,ihme2009}.\\
All the authors, from the fundamental theoretical work by Clavin and Siggia \cite{clavin1991} to the systematic study by Rajaram and Lieuwen \cite{rajaram2009}, agree on the shape of the combustion noise spectrum $S_{pp}$. In \cite{rajaram2009} it has been shown that the normalized combustion noise power spectra of different burners operating under different conditions collapse on top of each other, indicating a general scaling law (see \cref{fig:exp_Spp_engine}.a). The combustion noise spectrum features a maximum, and varies like $S_{pp}(f) \propto f^{2}$ on the left side, and like $f^{-r}$, with $2<r<3.4$, on the right side. The peak frequency of the combustion noise spectrum can be estimated making use of experimental relations such as the one proposed in \cite{shivashankara1975}, involving dimensions, flow properties and chemical quantities. Alternatively, it has been observed in \cite{rajaram2009} that the Strouhal number $St=f_{\mathrm{max}}{L_{\mathrm{F}}}/{U_{\mathrm{avg}}}$ is almost in any case close to 1, where $L_{\mathrm{F}}$ is the flame length and $U_\text{avg}$ the average velocity of the reactants mixture. Hence $f_{\mathrm{max}}\approx{U_{\mathrm{avg}}}/{L_{\mathrm{F}}}$, which is shown in the inset of \cref{fig:exp_Spp_engine}.a.\\
\\
\noindent{}As exemplified in \cref{fig:exp_Spp_engine}, the acoustic signature dramatically changes when the flame is placed within a combustion chamber. In \cref{fig:exp_Spp_engine}b, a single mode dominates the spectrum, with a sharp peak of frequency $f_0$, surrounded by several side peaks,  which correspond to the other thermoacoustic eigenmodes. One can conveniently express the acoustic pressure at a given location $x$ as
\begin{equation} 
\label{P_sum_eigenmodes}
p(x,t)=\sum_{i=1}^{\infty}\eta_{i}(t)\psi_i(x),
\end{equation}
where $\eta_i$ denotes the amplitude of the $i^{th}$ mode and $\psi_i$ the spatial distribution of the corresponding natural acoustic mode of the chamber.

This spatial projection leads to a set of coupled stochastic nonlinear differential equations for the modes $\boldsymbol{\eta}(t)=\left[\eta_1(t),\cdots,\eta_j(t),\cdots\right]^T$.
However, it is often possible to describe the dynamics of a single mode $j$ by neglecting the influence of other modes \cite{culick2006book}.
In this case, the mode amplitude $\eta_j$ is given by the nonlinear stochastic oscillator 
\begin{equation} 
\label{modeamp}
\ddot{\eta}_{j}+\omega^{2}_{j}{\eta}_{j}=g_{j}({\eta}_{j},\dot{\eta}_{j}) +\xi_{j},
\end{equation} 
where $\omega_j=2\pi f_j$ is the angular frequency of the $j^\text{th}$ natural acoustic mode, $\xi_{j}(t)$ is the additive stochastic forcing coming from turbulence-induced processes. The term $g_j$ is a non-linear function which includes, amongst others, the effects of acoustic damping mechanisms and coherent heat release rate fluctuations (this last contribution is coherent in the sense that it depends on the acoustic field). One can see in \cref{fig:network}.b a diagram depicting the coherent feedback $\widehat{q}_\text{a}$ from a flame located in a combustion chamber. At the same time the flame is also influenced by the turbulent flow. The resulting heat release fluctuation is the aforementioned  $\widehat{q}_\text{n}$, which was the only source in case of an open flame. The turbulence-induced flow perturbations exhibit a much smaller spatial correlation than the acoustic ones, which are correlated over the entire combustor. These quantities $\widehat{q}_\text{a}(\omega)$ and $\widehat{q}_\text{n}(\omega)$ can be measured in dedicated test rigs equipped with loudspeakers and microphones, as explained in e.g. \cite{paschereit2002}, and can be used afterwards in network models providing predictions of the system stability.\\
\subsection{Colored noise driven Van der Pol oscillator}
\label{VDP_model}
In the following, it is assumed that the non-linear function $g_j$ in \cref{modeamp} results from a linear acoustic damping and a non-linear flame feedback: $g({\eta},\dot{\eta})=\dot{q}_{\mathrm{a}} -\alpha\dot{\eta}$, where $\alpha$ is the damping constant, and subscripts are omitted from now on. 
The flame response is expanded up to the third order in acoustic amplitude, which is often sufficient to characterise supercritical thermoacoustic bifurcations \cite{lieuwen03,boujo2016}: ${q}_{\mathrm{a}}=\beta\eta-{\kappa}\eta^3/3$. This assumption yields the already presented Van der Pol oscillator equation:
\begin{equation} 
\label{VDP}
\ddot{\eta}+\omega_0^{2}{\eta}=[2\nu-\kappa\eta^{2}]\dot{\eta} +\xi(t)
 \end{equation}
where $\nu=(\beta -\alpha)/2$ is the linear growth rate.\\

Considering $\xi$ as a white noise, i.e. a delta correlated forcing, simplifies the modeling approach and has been used in most of the studies dealing with stochastically forced  thermoacoustic limit cycles (e.g., again \cite{lieuwen03,boujo2016}).\\
In the remainder of the paper the random forcing $\xi$ is assumed to result from the non-coherent heat release rate fluctuations $q_\text{n}$ only.  As a result, $S_{\xi\xi}$ follows the same power law as the combustion noise and is therefore proportional to $S_{pp}$  \cite{ihme2009, liu2015}.\\ 

\begin{figure}[t!]
\begin{center}
\includegraphics[width=0.5\columnwidth]{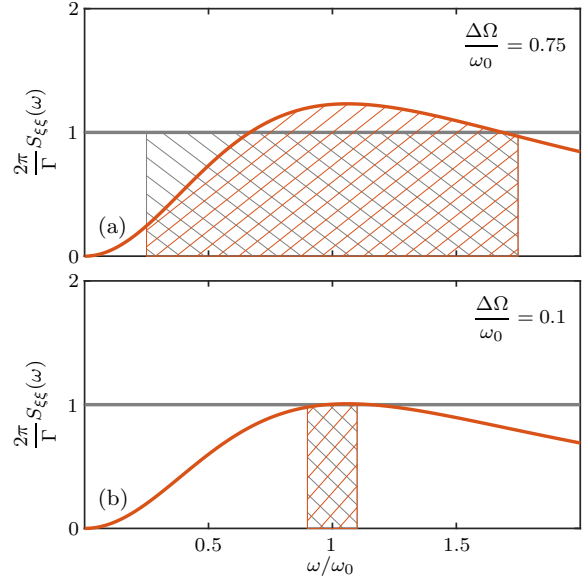}
\end{center}
\caption{Comparison between white noise (grey) and colored noise (red) power spectra, normalized by the white noise intensity, for different iso-power bandwidths $\Delta\Omega$. The power provided by the two types of noise is equal in the considered band (same area under the curve: note the linear scale). Note that $S_{\xi\xi}(\omega_0)=\Gamma_\text{e}/2\pi \neq \Gamma/2\pi$.\label{fig:isopower}} 
\end{figure}
\begin{figure*}
\begin{center}
\includegraphics[width=\textwidth]{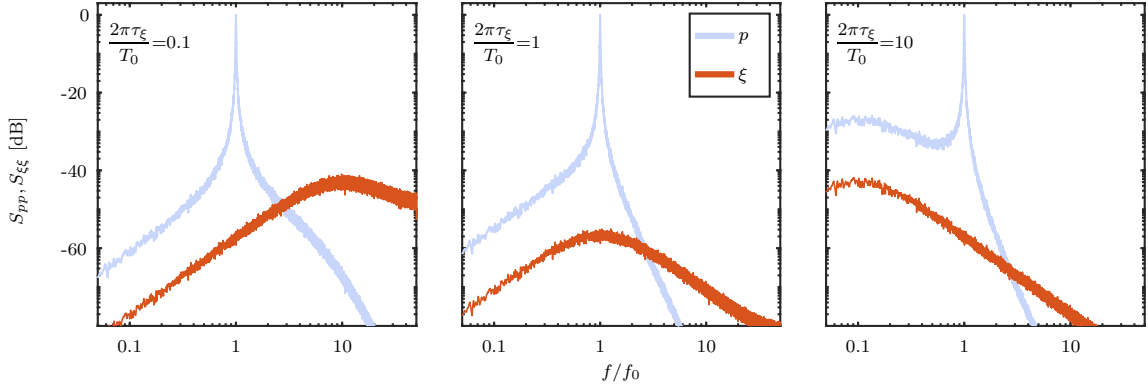}
\end{center}
\caption{Mutual position of noise and pressure spectrum maxima for three different adimensional noise correlation times $2\pi\tau_\xi/T_0$. Depending on the correlation time of the forcing noise, its spectrum maximum changes position accordingly (\cref{fmax_noise}). This modifies the response of the system, as can be observed on the output spectra. \label{fig:BP_fmax}}
\end{figure*}
\noindent{}In order to keep the problem tractable, $\xi$ is defined by 
\begin{equation}
\label{transfer_function_noise}
\widehat{\xi}(s)=H(s)\,\widehat{\zeta}(s)=\frac{\sqrt{\gamma}\tau^{2}s}{(1+\tau s)^{2}}\,\widehat{\zeta}(s),
\end{equation}
where   $\zeta$ is a unit-variance Gaussian white noise of intensity $\Gamma$, $\gamma$ is a constant used to adjust the power of the process $\xi$ and $\tau$ is its  characteristic time constant.
The resulting power spectrum is given by $|H|^2S_{\zeta\zeta}$:
\begin{equation}
\label{power_spectrum_noise}
S_{\xi\xi}(\omega)=\frac{\Gamma}{2\pi}\frac{\gamma\omega^{2}\tau^{4}}{(1+\omega^{2}\tau^{2})^{2}},
\end{equation}
that features a maximum at
\begin{equation}
\label{fmax_noise}
f_\text{max}=\frac{1}{2\pi\tau}.
\end{equation}
One can again define an ``effective colored noise intensity":
\begin{equation}
\label{Gamma_bp}
\Gamma_\text{e}=2\pi S_{\xi\xi}(\omega_0)=\Gamma\frac{\gamma\omega_0^{2}\tau^{4}}{(1+\omega_0^{2}\tau^{2})^{2}}.
\end{equation}
This model is a close approximation of actual experimental data, as shown in \cref{fig:exp_Spp_Sqq_open} (-$\Circle$-). This model is also close to others provided in literature, like in \cite{liu2015}, but, thanks to its simplicity, it allows for the analytical derivation that follows.\\
As done for the OU case, the colored noise power is equated to the one of a white noise of intensity $\Gamma$ in the band $[\omega_1;\omega_2]=\omega_0\pm\Delta\Omega$, which yields:
\begin{equation}
\label{intensity_noise}
\gamma=\frac{2(\omega_{2}-\omega_{1})}{\tau}\left(\text{atan}(\omega_{2}\tau)-\text{atan}(\omega_{1}\tau)-\frac{\omega_{2}\tau}{1+\omega_{2}^{2}\tau^{2}}+\frac{\omega_{1}\tau}{1+\omega_{1}^{2}\tau^{2}}\right)^{-1}.
\end{equation}
One can see in \cref{fig:isopower} how the parameter $\Delta\Omega$ affects the forcing noise power spectrum.\\
The characteristic time $\tau$ is related to the noise correlation time $\tau_\xi$, that can be obtained via the autocorrelation function of $\xi$,
\begin{equation}
\label{autocorrelation_function}
k_{\xi\xi}(t)=\Gamma\frac{\gamma\tau^{2}}{4\sqrt{2\pi}}(\tau-t)e^{-\frac{t}{\tau}},
\end{equation}
\begin{equation}
\label{autocorrelation_time}
\tau_\xi=\frac{1}{k_{\xi\xi}(0)}\int_{0}^{\infty}|k_{\xi\xi}(t)|dt=\frac{2\tau}{e},
\end{equation}
where $e=\exp{(1)} \simeq 2.718$.\\
The value of $\tau_\xi$ is related to the ``color'' of the noise: it determines where the maximum of the noise spectrum $f_\text{max}$ is located compared to the oscillator eigenfrequency $f_0$, affecting, as presented in \cref{fig:BP_fmax}, the response of the VDP. Focusing in a band around $f_0$, one can see how the oscillator is forced either by a source having the power increasing with frequency (``blue'' noise), or almost constant (close to a white noise), or decreasing (``pink'' noise). The resulting output $p$ is, accordingly, substantially different.
\begin{figure}[h!]
\begin{center}
{\includegraphics[width=0.5\columnwidth]{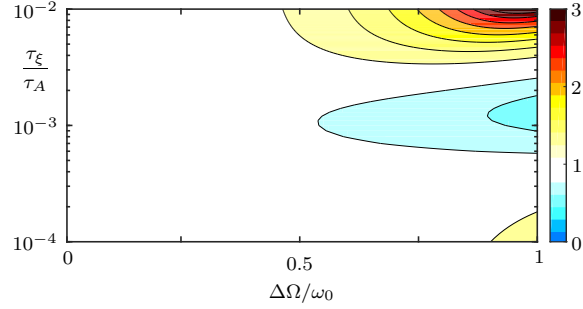}}
\end{center}
\caption{Map of the coefficient $\Gamma/\Gamma_\text{e}={(1+\omega_{0}^{2}\tau^{2})^{2}}/{\gamma\omega_{0}^{2}\tau^{4}}$. The closer this is to one, the closer the analytical expressions for $P_{\mathrm{c}}$ and $P_{\mathrm{w}}$ are. Note that the coefficient can be either greater than one (red scale) or smaller (blue scale).\label{fig:fact_map}}
\end{figure}
\begin{figure}[h!]
\begin{center}
\includegraphics[width=0.5\columnwidth]{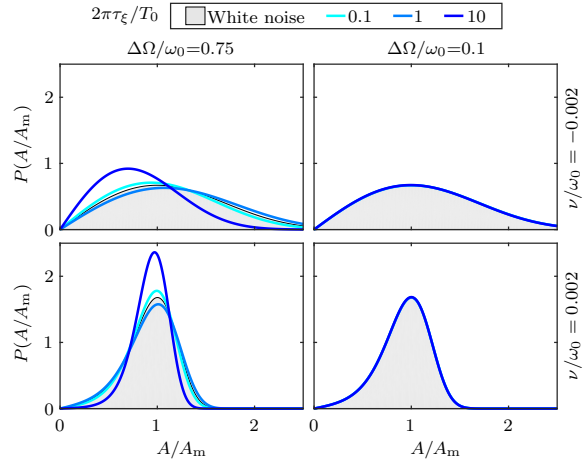}
\end{center}
\caption{Probability density function for two different linear growth rates $\nu$, two different iso-power semi-bandwidth $\Delta\Omega$, and three different adimensional correlation times $2\pi\tau_\xi/T_0$ of the driving noise (where $T_0=2\pi/\omega_0$ is the acoustic period). Solid lines are the PDFs for colored noise VDP (\cref{PAc}), shaded area the white noise driven PDF (\cref{PAw}) for the same parameters. The amplitude $A$ is given as relative to $A_{\mathrm{m}}$, the amplitude where $P_{\mathrm{w}}(A)$ is maximum.\label{fig:vdp_pdf}}
\end{figure}
%
\begin{figure*}[t!]
\begin{center}
\includegraphics[width=\textwidth]{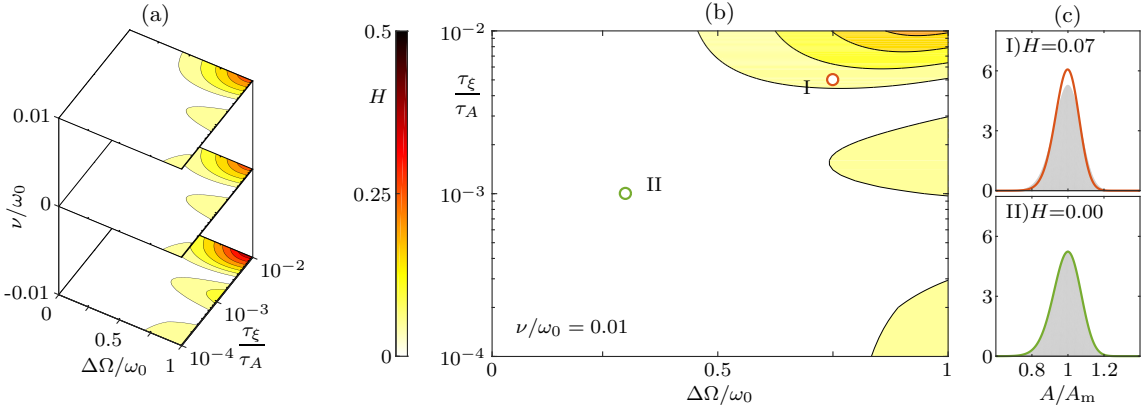}
\end{center}
\caption{Hellinger distance (\ref{H}), quantifying the difference between the PDFs of colored noise and white noise driven VDP oscillators. a) Different maps in the space $(\Delta\Omega,\tau_\xi,\nu)$. b) Detail of a linearly unstable case. c) PDFs for two linearly unstable points.\label{fig:H}}
\end{figure*}
\\
\noindent{}The VDP equation is again recast in amplitude-phase coordinates. In this case, this substitution is legitimate as, in most of the practical cases, the thermoacoustic systems satisfy the condition $\nu\ll\omega_0$. This means that the right hand side of \cref{modeamp} is much smaller than the left one and then $\eta(t)\approx A(t)\cos{[\omega t + \varphi(t)]}=A(t)\cos{\phi(t)}$. 
Adopting the colored noise model (\ref{power_spectrum_noise}) for $\xi$, deterministic and stochastic averaging yields the stochastic differential equation:
\begin{equation}
\dot{A}=A\left(\nu-\frac{\kappa}{8}A^{2}\right)+\frac{\Gamma_\text{e}}{4\omega_{0}^{2}A}+\mu(t), \hspace{5mm} \langle\mu\mu_\tau\rangle=\frac{\pi S_{\xi\xi}(\omega_0)\delta(\tau)}{\omega_0^2}=\frac{\Gamma_\text{e}}{2\omega_0^2}\delta(\tau),
\label{A_dot_vdp_avg} 
\end{equation}
with $\Gamma_\text{e}$ given by \cref{Gamma_bp}.
\noindent{}Again, the averaging method is valid if  the correlation times are such that $\tau_{\xi}\ll\tau_A$ \cite{stratonovich1967}. This is generally verified for practical cases. The amplitude correlation time is related to the growth rate by $\tau_A \simeq \pi/|\nu|$ \cite{noiray16}. Taking $\nu=10$ rad/s, for instance, $\tau_A=314$ ms, while the noise correlation time $\tau_\xi=2\tau/e\approx1/e\pi f_{\mathrm{max}}$ is generally smaller than 1 ms ($f_{\mathrm{max}} \geq 50$ Hz, see \cref{fig:exp_Spp_engine}.a).\\
The stationary probability distribution for the amplitude of the bandpass noise driven VDP oscillator is then:
\begin{equation}
P_{\text{c}}(A)=\mathcal{N}_\text{c}A\exp{\left[\frac{4\omega_{0}^{2}}{\Gamma_\text{e}}\left(\frac{\nu A^{2}}{2}-\frac{\kappa A^{4}}{32}\right)\right]},
\label{PAc}
\end{equation}
where $\mathcal{N}_{\mathrm{c}}$ is the normalization constant to have $\int_{0}^{\infty}P_\text{c}(A)dA=1$.\\
Like for the OU case, \cref{A_dot_vdp_avg,PAc} have the same structure as their white noise driven system counterparts \cref{Aw_dot_vdp_avg,PAw}, with the effective colored noise intensity $\Gamma_\text{e}$ (\cref{Gamma_bp}) replacing the white noise intensity $\Gamma$. Therefore, $P_{\mathrm{c}}$ and $P_{\mathrm{w}}$ differ only for the factor $\Gamma/\Gamma_\text{e}$ in the exponential. In \cref{fig:fact_map} one can see a map of this factor, as function of the iso-power bandwidth $\Delta\Omega$ and of the source noise correlation time $\tau_\xi$.
\\
\noindent{}Comparing this map with the one for the OU noise (\cref{fig:OU_fact_map}), one can notice how $\Gamma$ and $\Gamma_\text{e}$ might differ whatever the correlation time. This is due to the fact that this type of noise does not converge to a white one for short $\tau_\xi$. Another difference is that this coefficient can be lower than one.\\
\noindent{}In line with this map, $P_{\mathrm{c}}$ and $P_{\mathrm{w}}$ can show significant differences, as depicted in \cref{fig:vdp_pdf}.
To compare quantitatively $P_\text{c}$ and $P_\text{w}$, the Hellinger distance is plotted in \cref{fig:H}. As for the OU noise, for small $\Delta\Omega$, $H$ tends to 0. However in this case, for large $\Delta\Omega$, $H$ is large whatever the correlation time of the noise source.\\
\section{Thermoacoustic instabilities: parameter identification}
\label{Results}
\noindent In this section, the white noise approximation is assessed in the context of  parameter identification. As discussed before, the dynamics of a thermoacoustic mode can be seen as a SISO system. Although the output, represented by pressure oscillation, is easily accessible via experimental measures, the input, resulting from turbulence, is not known. Therefore, this system necessitates output-only parameter identification methods.
\subsection{Assessment of the white noise approximation}
\label{Validity of the white noise approximation}
%
%
\begin{figure}[h!]
\begin{center}
\includegraphics[width=0.5\columnwidth]{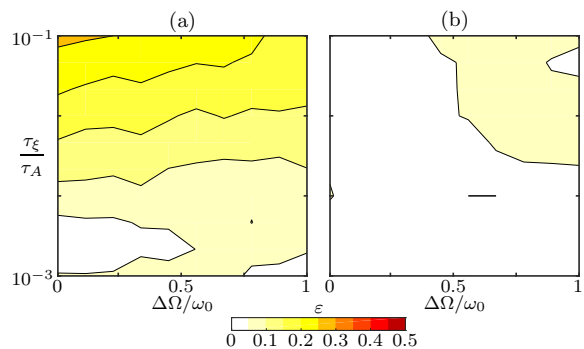}
\end{center}
\caption{Map of band-pass colored noise driven oscillator identification, for different noise correlation time $\tau_\xi$ and iso-power semi-bandwidth $\Delta\Omega$. a) Identification using the unfiltered data. b) Identification using signals filtered in the band $[f_0(1\pm0.5)]$. The identification result is given as relative error: $\varepsilon=|\nu_\text{t}-\nu_\text{id}|/\nu_\text{t}$.\label{fig:BP_ID_map}}
\end{figure}
\begin{figure*}[th!]
\begin{center}
\includegraphics[width=\textwidth]{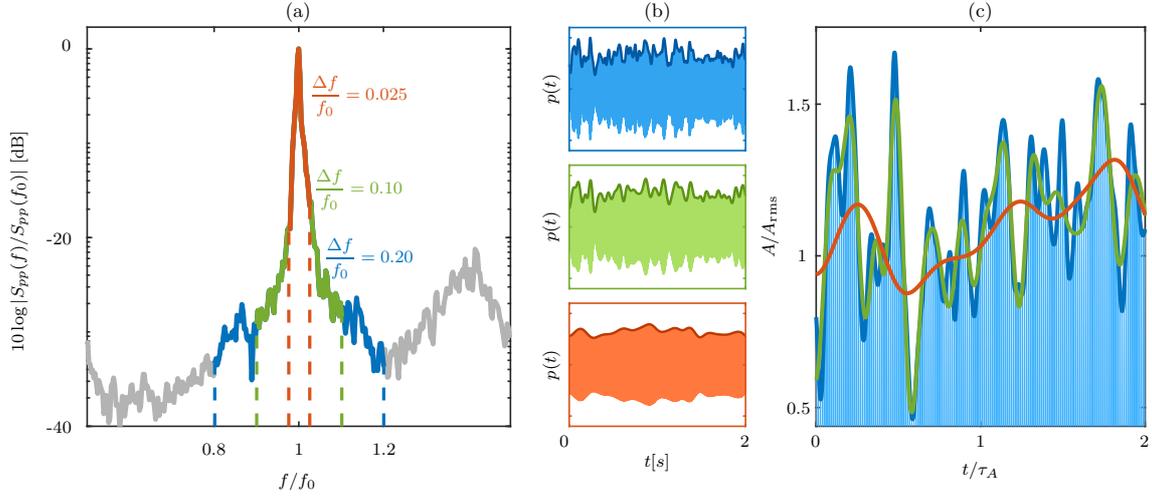}
\end{center}
\caption{Effects of three different filter bandwidths on the analysis of combustor pressure experimental data. a) Acoustic pressure spectrum and filters bands. b) Time traces resulting from the three different filtering. c) Detail of the envelopes over a time span of two $\tau_A$.\label{fig:filt_w}}
\end{figure*}
\noindent{}Following the same procedure as in \cref{OU2}, 100 test cases with fixed oscillator parameter $\nu=\nu_\text{t}$ and $\kappa=\kappa_\text{t}$, but different noise parameters $\tau_\xi$ and $\Delta\Omega$, are analysed to ensure that the identification methods relying on the white noise assumption are not biased by the actual noise spectrum and autocorrelation. The relative error $\varepsilon=|\nu_\text{t}-\nu_\text{id}|/\nu_\text{t}$ on the estimated oscillator linear growth rate $\nu_\text{id}$ is presented in \cref{fig:BP_ID_map}.\\
\noindent{}Like for the OU noise case, the identification might fail if the unfiltered data are used (left panel). It is interesting to notice that, compared to the OU case, the error is generally less severe. This is due to the spectral distribution of the bandpass noise, rapidly decaying in power at high and low frequencies. Another peculiar aspect is the distribution of the errors in the map. While for the OU noise low $\tau_\xi$ means a quasi-white noise forcing and, therefore, a small identification error, here short $\tau_\xi$ corresponds to a blue noise forcing.\\
As in the OU case, filtering the data prior to parameter identification improves the identification results (right panel of \cref{fig:BP_ID_map}). This is, again, due to a more accurate approximation of the real forcing spectrum with a white one in the considered frequency range. \\
The bandwidth of the filter adopted in the pre-processing of the data has to be chosen with care in order not to discard  essential amplitude dynamics. In addition, practical acoustic spectra often feature neighboring peaks around the main one, due to the coexistence of several thermoacoustic modes in the combustor. This is a further constraint when one analyses experimental data and performs single-mode output-only parameter identification. These two aspects are covered in the following.  
\subsection{Effect of data preprocessing on parameter identification}
\label{Filter size for system identification}
\begin{figure*}[h!]
\begin{center}
\includegraphics[width=\textwidth]{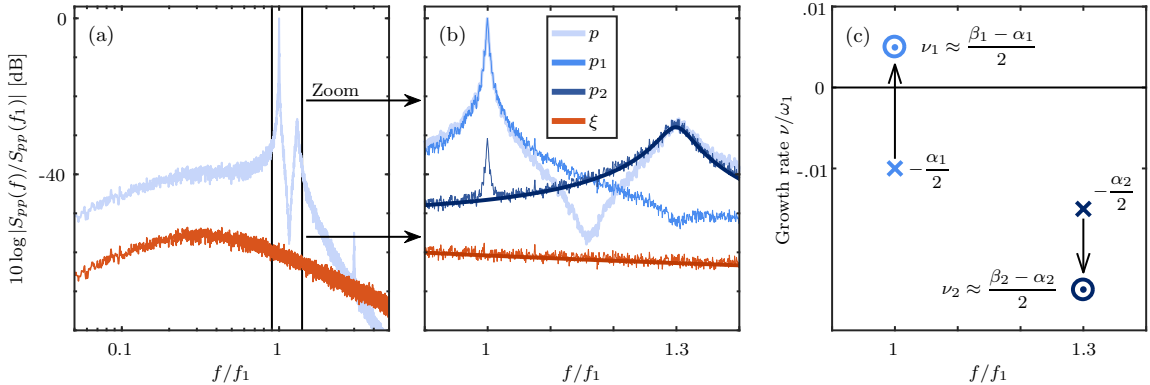}
\end{center}
\caption{Double-oscillator simulation results. The model is made of two non-linear oscillators linearly coupled, and is fed by colored noise. Oscillator \#1 is linearly unstable, oscillator \#2 is stable. Values adopted for the parameters (refer to \cref{2VDP_eq}): $\omega_2=1.3\omega_1$, $\alpha_1/\omega_1=0.02$, $\beta_1/\omega_1=0.03$, $\alpha_2/\omega_1=0.03$, $\beta_2/\omega_1=-0.02$, $\kappa_1/\omega_1=\kappa_2/\omega_1=0.015$. The colored noise parameters (refer to \cref{power_spectrum_noise}) are: $f_\text{max}/f_1=0.2$, $\Delta\Omega/\omega_1=0.5$, $\Gamma/4\omega_1^2=1$.
a) Overview of the total pressure and forcing noise spectra. b) The spectral SPL of total output, single oscillators outputs $p_1$ and $p_2$, and forcing noise, in the frequency band that encloses the two oscillators' natural frequencies. c) The poles of the linearised coupled system move due to the feedback action, that can either decrease or increase the stability margin of each mode, or even fully destabilize a mode.\label{fig:vdp2}}
\end{figure*}
\begin{figure}[h!]
\begin{center}
\includegraphics[width=0.5\columnwidth]{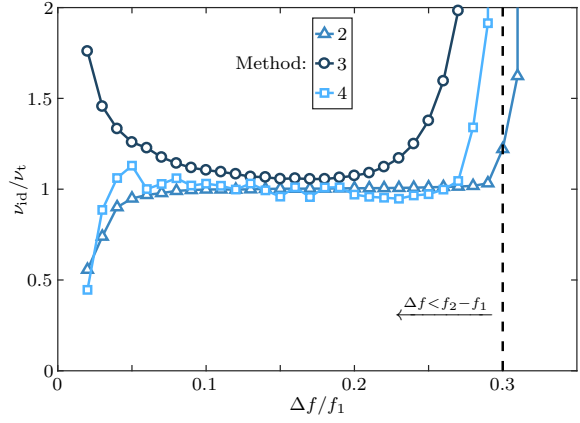}
\end{center}
\caption{Identified growth rate against the filter semi-bandwidth. The source signal is obtained via a Simulink\textsuperscript{\textregistered} simulation of a double VDP oscillator, of known parameter (e.g. $\nu_{\text{t}}$ is the true growth rate). The identification is performed on mode \#1, of eigenfrequency $f_1$, while another mode of eigenfrequency $f_2$ is in place. Three identification methods of \cite{noiray13} are used, respectively based on: the power spectral density of the amplitude (-$\bigtriangleup$-), the probability density function of the amplitude (-$\Circle$-) and the coefficients of the Fokker-Planck equation (-$\square$-).\label{fig:nu_id}}
\end{figure}
\noindent A typical combustor acoustic pressure spectrum features several peaks (\cref{fig:exp_Spp_engine}.b). The different modes acting in the domain are mutually coupled, each one influencing the response of the others. However, if the neighboring peaks are not too close, one can analyse one mode at a time, isolating its dynamic from that of the other modes. This is easily done by bandpass filtering the data and simplifies the system identification, since neither the parameters of neighboring modes, nor the coupling coefficients have to be taken into account.\\
\Cref{fig:filt_w} shows a typical situation and the effects of a different filter bandwidth. A wider portion of this spectrum has already been shown in \cref{fig:exp_Spp_engine}.b. This experimental spectrum features a strong peak, corresponding to the dominant mode eigenfrequency, surrounded by two others small peaks. In order to identify the mode parameters accurately, removing the other modes effect, the signal is filtered around the main peak, i.e. in the band $[f_0-\Delta f; f_0+\Delta f]$. The maximum bandwidth is the one that discards neighboring peaks while keeping the main peak and its tails ($\Delta f/f_0=0.20$ in this case). One could also choose narrower bands ($\Delta f/f_0=0.10$ or $\Delta f/f_0=0.025$ in this example), obtaining different resulting time signals. Looking at the central panels of \cref{fig:filt_w}, one can see that in the first case (green), the dynamics on time scales comparable to the amplitude correlation time $\tau_A=\pi/|\nu|$ is preserved: compared to the widest filter (blue), only high-frequency amplitude oscillations are lost. This means that the essential dynamics are unaffected. In the second case (red), the general trend is followed, but too much information has been lost to reliably identify the parameters $\nu$ and $\kappa$.\\
In the following, a ``toy model" of two coupled oscillators driven by colored noise is used to illustrate this issue:
 \begin{equation} 
\label{2VDP_eq}
\begin{cases}
\ddot{\eta}_1+\alpha_1\dot{\eta}_1+\omega_1^2\eta_1=[\beta_1-\kappa_1\eta_1^2]\dot{\eta_1} + [\beta_2-\kappa_2\eta_2^2]\dot{\eta_2} + \xi \\
\ddot{\eta}_2+\alpha_2\dot{\eta}_2+\omega_2^2\eta_2=[\beta_1-\kappa_1\eta_1^2]\dot{\eta_1} + [\beta_2-\kappa_2\eta_2^2]\dot{\eta_2} + \xi \\
p=p_1+p_2=\psi_1\eta_1+\psi_2\eta_2.
\end{cases}
\end{equation}
The total output $p$, which is the sum of the outputs of the two oscillators $\eta_{1}$ and $\eta_{2}$ weighted by $\psi_1$ and $\psi_2$, features a spectrum, plotted in \cref{fig:vdp2}, that is similar to the experimental pressure spectrum shown in \cref{fig:exp_Spp_engine}. This figure also highlights what is hidden behind a single-mode approximation.\\
\noindent{}Note the difference between the spectra of $p_2$ without coupling (theoretical, thick blue) and with coupling (numerical, thin blue), especially for $f=f_1$. This difference appears because the oscillators are coupled and the linearly unstable oscillator \#1, characterized by a limit-cycle at $f_1=\omega_1/2\pi$, is forcing oscillator \#2, having eigenfrequency $f_2=1.3f_1=\omega_2/2\pi$. At the same time, the linearly stable mode (oscillator \#2) contributes to $S_{pp}$ around the eigenfrequency $f_1$ of the unstable mode (oscillator \#1). 
Therefore, the response of the system at $f=f_1$ is not due to the oscillator \#1 only. 
However, if the two peaks are distant enough and one is stronger than the other, these mutual contributions are negligible, compared to the direct output of the oscillator \#1 at its natural frequency (more than 20 dB of difference in this example). Restricting the discussion to this case, one can adopt the aforementioned single-mode approximation, and attempt a parameter identification on one mode at a time.\\
\noindent{}To test the sensibility of the identification results to the filter bandwidth, the output signal is filtered with different bandwidth around the first eigenfrequency $f_1$. The aim is to extract the linear growth rate of the first unstable oscillator, which has the true value $\nu_{\text{t}} \approx (\beta_1-\alpha_1)/2$. For this purpose, three different methods of \cite{noiray13} are used. The results are presented in \cref{fig:nu_id}. One can observe that, whatever the adopted identification method, for too narrow filter bandwidth, the identified growth rate is far from the true one, whereas it converges to $\nu_\text{t}$ for large enough windows. On the other hand, when the filter is too wide, the effect of the neighboring mode starts to bias the identification.
Therefore, when one analyses experimental data around a frequency of interest, there exist, for the filter bandwidth: i) a lower limit, given by the need not to alter the amplitude statistics,
ii) an upper limit, given by the distance from the neighboring peaks.
These constraints have to be satisfied in parallel with the one regarding the validity of the white noise approximation (\cref{Validity of the white noise approximation}). However, in most of the practical cases, neighboring peaks are close and the maximum filter bandwidth to satisfy condition ii) is narrow enough that the effect of noise color can be safely neglected.\\
On the other hand, it is clear that any identification attempt on a mode that is both highly unstable and very close to another mode will fail because the filter to adopt to isolate one mode dynamics would be so narrow that condition i) is not fulfilled. In this situation a two-mode model would be required for parameter identification. As already suggested, it is advisable to iterate the parameter identification varying the applied filter bandwidth: one can be confident on the result if a plateau is observed.
\section{Conclusion}
\noindent In this work, the effects of the color of a stochastic excitation driving a Van der Pol Oscillator has been investigated. 
First, an Ornstein-Uhlenbeck process has been considered as the driving source. Then, a noise model, mimicking the stochastic forcing exerted by turbulence in thermoacoustic systems, has been used.
It has been shown that in both cases the envelope statistics is the same as the one obtained with a white noise forcing, provided that an equivalent effective noise intensity is considered. 
Then, the approximation of a colored noise by a white one has been assessed in the context of data analysis and parameter identification.
The main conclusion is that one can reliably identify the linear growth rate and saturation constant by band-pass filtering the data around the oscillator eigenfrequency.\\
This result is valid regardless of the parameters values and nature of the forcing noise. This fact consolidates the output-only parameter identification methods proposed in \cite{noiray13}, because in real cases it might be impossible to determine the spectral distribution of the forcing noise.\\
\section*{Acknowledgement}
\noindent{}This research is supported by the Swiss National Science Foundation under Grant 160579.
\bibliographystyle{apsrev4-1}
\bibliography{./Bonciolini_Colored_Noise} 
\end{document}